\documentclass[twocolumn,prb]{revtex4}
\usepackage{graphicx}
\usepackage{dcolumn}
\usepackage{bm}
\usepackage{amsmath}
\usepackage{amsfonts}
\usepackage{feynmp}
\usepackage{psfrag}

\newcommand{\ket}[1]{|#1\rangle}

\newcommand{\sumk}[1]{\sum_{\mathbf{#1}}}
\newcommand{\Eq}[1]{Eq. (\ref{#1})}
\newcommand{\ca}[2]{c_{#1,\mathbf{#2}}}
\newcommand{\cc}[2]{c^{\dagger}_{#1, \mathbf{#2}}}
\newcommand{\ba}[2][]{b^{#1}_{\mathbf{#2}}}
\newcommand{\bc}[2][]{b^{\dagger\,#1}_{\mathbf{#2}}}
\newcommand{\da}[2]{d_{\mathbf{#1},#2}}
\newcommand{\dc}[2]{d^{\dagger}_{\mathbf{#1},#2}}
\newcommand{\ra}[1]{r_{\mathbf{#1}}}
\newcommand{\rc}[1]{r^{\dagger}_{\mathbf{#1}}}
\newcommand{\comm}[2]{\lbrack #1,#2\rbrack}
\newcommand{\vh}{\hat{v}}
\newcommand{\vhd}{\hat{v}^\dagger}
\newcommand{\Mc}{{\mathcal H}}

\begin{document}
\title{Quantum theory of intersubband polarons}

 \author{Simone \surname{De Liberato}}
\author{Cristiano Ciuti}
\affiliation{Laboratoire Mat\'eriaux et Ph\'enom\`enes Quantiques, Universit\'e Paris  Diderot-Paris 7 and CNRS, UMR 7162, 75013 Paris, France} 
 
 \begin{abstract}
We present a microscopic quantum theory of intersubband polarons, quasiparticles originated from the coupling between intersubband transitions and longitudinal optical phonons. To this aim we develop a second quantized theory taking into account both the Fr\"ohlich interaction between phonons and intersubband transitions and the Coulomb interaction between the intersubband transitions themselves.
Our results show that the coupling between the phonons and the intersubband transitions is extremely intense, thanks both to the collective nature of the intersubband excitations and to the natural tight confinement of optical phonons. 
Not only the coupling is strong enough to spectroscopically resolve the resonant splitting between the modes (strong coupling regime), but it can become comparable to
the bare frequency of the excitations (ultrastrong coupling regime).
We thus predict the possibility to exploit intersubband polarons both for applied optoelectronic research, where a precise control of the phonon resonances is needed, and also 
to observe fundamental quantum vacuum physics, typical of the ultrastrong coupling regime.
\end{abstract}

 \maketitle
 \section{Introduction}
 
The theory of polarons, the quasiparticles describing electrons in a polarizable medium, dates back to the early days of quantum theory  \cite{Landau33}, and it has been an active field of research ever since \cite{Devreese96}.

In this paper we will develop a microscopic theory of intersubband polarons, that is, a theory of intersubband transitions coupled to longitudinal optical (LO) phonons in semiconductor quantum wells.

The coupling between intersubband transitions and LO-phonons is relevant for a number of optoelectronic applications, as it determines the lifetime of carriers in excited subbands \cite{Ferreira89}. In particular a precise knowledge of LO-phonons intersubband scattering rates is important in the engineering of heterostructures for quantum cascade lasers \cite{Faist94}.
Normally optoelectronic devices are designed to avoid being in resonance with optical phonon transitions, due to the high absorption between transverse and longitudinal optical phonon frequencies (Restrahlen band). A notable exception is provided by quantum cascade lasers operating near such optical resonances\cite{Colombelli01,Castellano11}, in which instead the transitions between different subbands are almost resonant with LO-phonon modes.
 
Even if the coupling between intersubband transitions and LO-phonons in semiconductor quantum wells has indeed received some attention \cite{Butscher04,Butscher06,Cao06} 
and intersubband polaron resonances have been clearly and unambiguously observed \cite{Liu03}, to the best of our knowledge, there is no microscopic theory of such excitations, as the spectra of intersubband polarons are normally calculated with indirect methods. 
While such methods allow to calculate, at least qualitatively, the polaron dispersions, missing a microscopic description makes it difficult to study more complex phenomena as nonequilibrium physics,  quantum vacuum effects or quantum phase transitions.

Using a second quantization formalism, we will reduce the full electron-phonon Hamiltonian to a quadratic, bosonic form, from which we will then calculate the polaron dispersions.
In order to accomplish this task, we show that Coulomb interaction between electrons in conduction subbands naturally
separates into a dominant and a perturbative part, accordingly to the number of electrons that can participate to each transition.

Moreover we will show how, thanks to the tight confinement of LO-phonons, the coupling between intersubband transitions and phonons can easily be in the ultrastrong coupling regime, a regime characterized by a coupling strength comparable to the bare frequency of the excitations \cite{Ciuti05}. Such fact can have interesting observable consequences, as a whole new range of physics is {\it a priori} observable in this regime:
spectral deformations  \cite{Anappara09,Todorov10,Hagenmuller10}, quantum vacuum emission phenomena \cite{Kardar99,DeLiberato07,DeLiberato09}, electroluminescence enhancement \cite{DeLiberato08,DeLiberato09c} and even quantum phase transitions \cite{Emary03,Lambert04,Nataf10b}. 

This article is organized as follows: in Sec. \ref{Theory} we will develop the general theory of the coupling between intersubband transitions and LO-phonons, that we will then apply in Sec. \ref{NumRes} to the case of GaAs quantum wells, showing how the ultrastrong coupling regime can be reached even with such relatively weakly polar material. In Sec. \ref{Appendix} we will compare the calculated dispersions  with the ones obtained using an homogeneous dielectric function approach.
Finally a few considerations on the impact of our results and on possible future developments will be drawn in Sec. \ref{Con}.

 \section{Theoretical framework}
 
 \label{Theory}

 \subsection{Superradiant excitations}

In 1954 Dicke \cite{Dicke54} noticed that a set of coherently excited identical dipoles relaxes radiatively much faster than a single, isolated one. This is due to  the phenomenon of superradiance:  $N$ identical dipoles behave as a single collective dipole $\sqrt{N}$ times bigger.  

The concept of superradiance has been thoroughly applied to the study of intersubband polaritons \cite{Dini03,Anappara07,Sapienza07,Sapienza08} in microcavity embedded quantum wells. 
In such systems the light couples to a collective electronic excitation  and, as expected \cite{Ciuti05}, the strength of the coupling between light and matter is proportional to the square root of the number of electrons involved.

\begin{figure}[t!]
\begin{center}
\includegraphics[width=8.5cm]{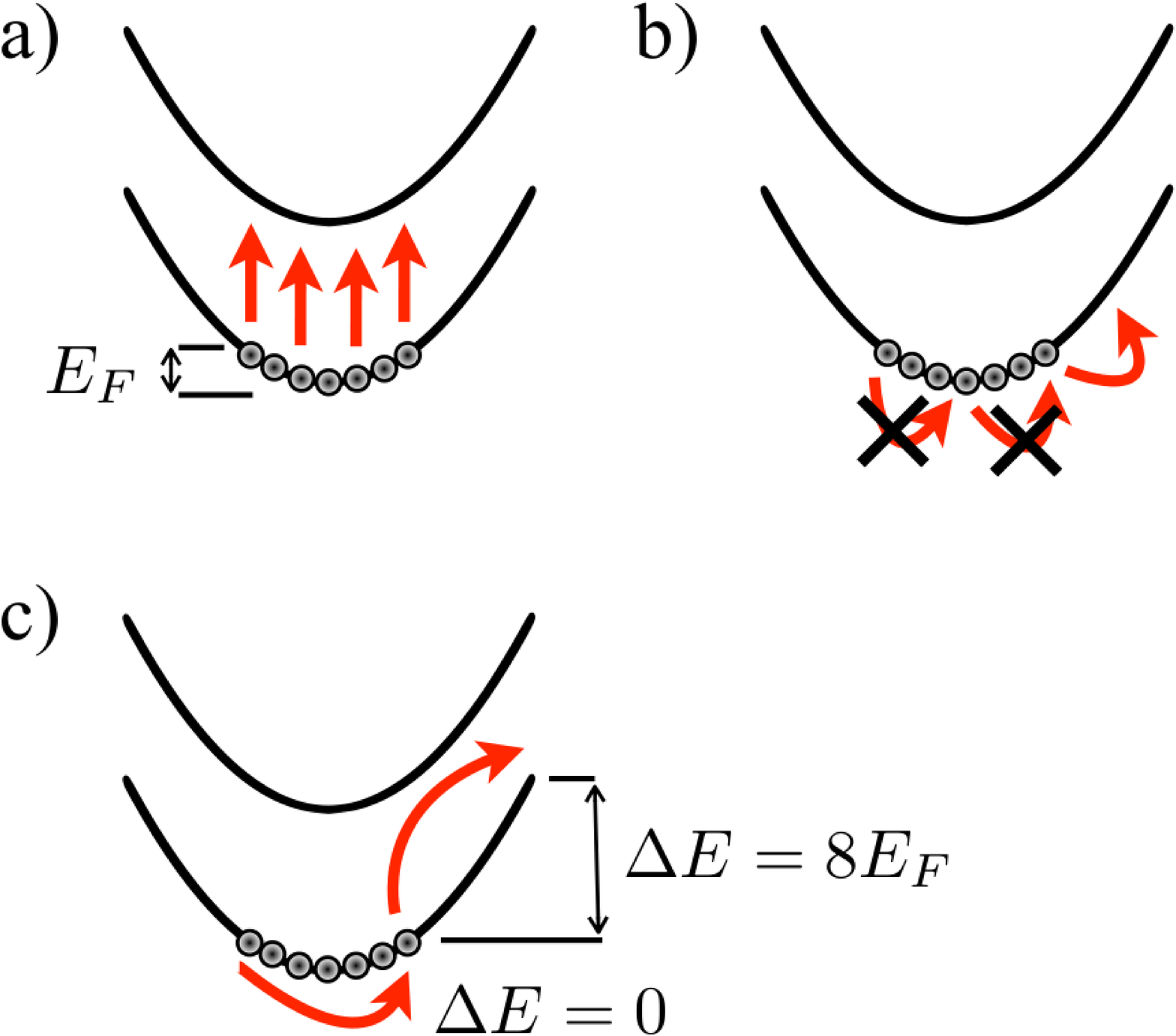}
\caption{ \label{Cohex} a) An example of collective transition: an intersubband transition with small transferred momentum. All the electrons can undergo a transition resonant approximately at the same energy.
 b) An example of non-collective transition: an intrasubband transition with transferred momentum much smaller than the fermi wavevector $\hbar k_F$. The majority of electrons, being Pauli blocked, do not participate to the process.  c) Another example of non-collective transition: an intrasubband transition with transferred momentum of the order of $2\hbar k_F$, that is the minimum to allow all the electrons to undergo the transition avoiding Pauli blocking. 
 Anyway there is a large energy spread between the different single electron transitions. 
For an infinite potential well, electrons on the two opposite borders of the Fermi sea, with initial momenta $\pm\hbar k_F$ parallel to the transferred momentum, have initially the same energy 
 $\frac{\hbar^2 k_F^2}{2m^*}=E_F$, where $m^*$ is the electron effective mass and $E_F$ the Fermi energy. After the transition they will end up with  final momenta $\hbar k_F$ and $3\hbar k_F$, corresponding to final energies $\frac{\hbar^2 k_F^2}{2m^*}=E_F$ and $\frac{9\hbar^2 k_F^2}{2m^*}=9E_F$ respectively. 
This implies that, even if the transition is not blocked, only a small fraction of the single electron transitions can be resonant at the same time. In both cases, given that only few electrons can participate to the collective transition, the superradiant enhancement factor will be small.}
\end{center}
\end{figure}

We will study the coupling of intersubband transitions with longitudinal optical phonons, considering also the role of Coulomb electron-electron interaction. 
Such couplings are extremely rich and, in order to limit the complexity of our investigation, we will need to determine which scattering channels are dominant  and which are negligible.
In general, if $N$ electrons undergo a certain transition in a coherent way, the strength of the coupling is enhanced by a factor $\sqrt{N}$. Transitions involving a macroscopic number of electrons will thus be strongly enhanced and, for this reason, they will be treated exactly within an Hamiltonian formalism, while the others (involving only few electrons ) will be treated perturbatively (or  ignored altogether). The degree of collective enhancement of a scattering process will be evaluated looking at the number of electrons that can undergo the transition given fixed amounts of transferred impulsion and energy.
In Fig. \ref{Cohex} we show a few illustrative examples of collective and non-collective transitions.

In the case of the Coulomb interaction, as we will see, only the intersubband terms, responsible for the depolarization shift, are collective. We will thus treat them exactly in the Hamiltonian, while considering all the other terms in a RPA linear response approach.

 \subsection{Free fields}
 
We will consider a symmetric quantum well of length $L_{QW}$ in a bulk of height $L_{BK}$,  $S$ will be the surface of the sample. For the moment, we will limit ourselves to the case of a single quantum well, the general case of multiple wells will be addressed later in this Section.
 
The quantum well is supposed to be doped in such a way that its Fermi level is between the first and the second conduction subbands, separated between them by the intersubband gap energy $\hbar\omega_{12}$.
 
We will develop our theory using a  zero temperature formalism ($T=0$), anyway
our results will remain quantitatively accurate while the thermal population of the second subband remains negligeable.
Depending on material parameters and doping level, this could imply the necessity to perform experiments in different kinds of cryogenic environments.

Electron states will be indexed by the subband index $j$ and by the value of the in-plane wavevector $\mathbf{k}$. Their wavefunctions will be given by 
\begin{eqnarray}
\label{electronicwf}
\psi_{j,\mathbf{k}}(\mathbf{\rho},z)=\chi_j(z)\frac{e^{i\mathbf{k\rho}}}{\sqrt{S}},\quad j=1,2,
\end{eqnarray}
where, for simplicity, we will choose $\chi_j(z)$ to be real and, due to the symmetry of the quantum well, the $\chi_j(z)$ have well defined and opposite symmetry.  
Wavefunctions in \Eq{electronicwf} are chosen as basis for second quantization, the creation operator for an electron in the state described by \Eq{electronicwf} will be denoted as $\cc{j}{k}$. The free Hamiltonian of the electron gas in the two considered subbands thus reads

\begin{eqnarray}
\label{Hfreec}
H_{el}=\sum_{j=\{1,2\},\mathbf{k}}  \hbar \omega_{j}(\mathbf{k})\cc{j}{k}\ca{j}{k},
\end{eqnarray}
where $\omega_1(\mathbf{k})=\frac{\hbar^2k^2}{2m^*}$ and $\omega_2(\mathbf{k})=\omega_1(\mathbf{k})+\omega_{12}$.
In \Eq{Hfreec}, as well as in the rest of this paper, we will omit the electron spin index. This is justified by the fact that all interactions we consider are spin conserving.
Given that we will consider only in-plane wavevector exchanges $\mathbf{q}$ much smaller than the typical electron wavevector $\mathbf{k}$, we can make the approximation  $\omega_j(\mathbf{k+q})\simeq \omega_j(\mathbf{k})$, and introduce
the operators describing intersubband transitions with a well defined and dispersionless energy $\hbar\omega_{12}$ \cite{Ciuti05}
\begin{eqnarray}
\label{bdef}
\bc{q}&=&\frac{1}{\sqrt{N}}\sumk{k}\cc{2}{k+q}\ca{1}{k},\\
\ba{q}&=&\frac{1}{\sqrt{N}}\sumk{k}\cc{1}{k}\ca{2}{k+q}\nonumber,
\end{eqnarray}
where $N$ is the number of electrons in the quantum well.
In the dilute regime, that is if the number of excitations is much smaller than $N$,
the $\bc{q}$ operators are bosonic 
\begin{eqnarray}
\label{bos}
\lbrack\ba{q},\bc{q'}\rbrack 
\simeq\delta_{\mathbf{q,q'}}.
\end{eqnarray}
At higher excitation densities, that are out of the scope of the present work, saturation effects start to appear and corrections to \Eq{bos} have to be taken into account. 
For a detailed analysis of nonbosonicity effects in intersubband transitions we invite the interested readers to refer to 
Ref. [\onlinecite{DeLiberato09b}].

Using \Eq{bdef} we can rewrite the Hamiltonian of the free electron gas in \Eq{Hfreec} in terms of bosonic intersubband excitations
\begin{eqnarray}
\label{Hfreeb}
H_{el}=\sum_{\mathbf{q}} \hbar\omega_{12}\bc{q}\ba{q}.
\end{eqnarray}

In this work we are interested in the resonant case in which $\omega_{12}$ is equal, or close, to the LO-phonon frequency $\omega_{LO}$. 
We can thus neglect confinement effects on the phonons and consider bulk values for their frequencies \cite{Arora87,Stroscio01}.
We will thus describe LO-phonons by means of the three dimensional boson operators 
$\da{q}{q_z}$
\begin{eqnarray}
\lbrack\da{q}{q_z},\dc{q'}{q_z'}\rbrack=\delta_{\mathbf{q,q'}}\delta_{q_z,q_z'},
\end{eqnarray}
indexed by their in-plane and out-of-plane wavevectors.  
While we know that LO-phonon modes are confined inside the quantum well, we do not need to impose this constraint in the mode definition
because, as we will see, intersubband transitions end up coupling with linear superpositions of phonon modes that are anyway confined inside the quantum well.
We will consider only the case of one single longitudinal optical branch, the expansion to the case of multiple branches not presenting any fundamental difficulty.
 
Moreover, we are interested only in phonons with small in-plane wavevectors (in order to couple with coherent intersubband excitations), we can thus ignore phonon dispersion and write the free phonon Hamiltonian as
\begin{eqnarray}
\label{Hfreed}
H_{ph}=\sum_{\mathbf{q},q_z}\hbar \omega_{LO}\dc{q}{q_z}\da{q}{q_z}.
\end{eqnarray}

\subsection{Electron phonon interaction}

Interaction between electrons and LO-phonons can be described using the Fr\"ohlich Hamiltonian \cite{Frohlich54}
\begin{eqnarray}
\label{HFFQ}
H_{Fr}=\sqrt{\frac{\hbar \omega_{LO}e^2}{2\epsilon_0\epsilon_{\rho}SL_{BK}}} \sum_{\mathbf{q},q_z} \frac{e^{-i(\mathbf{q}\mathbf{\rho}+q_z z)}}{\sqrt{q^2+q_z^2}} \dc{q}{q_z}+h.c.,
\end{eqnarray}
where 
\begin{eqnarray}
\label{epsdef}
\frac{1}{\epsilon_{\rho}}=\frac{1}{\epsilon_{\infty}}-\frac{1}{\epsilon_{s}},
\end{eqnarray}
where $\epsilon_{s}$ and $\epsilon_{\infty}$ are respectively  the static and high frequency dielectric constants \cite{Stroscio01}.

The Hamiltonian in \Eq{HFFQ} can be written in second quantization (neglecting incoherent intrasubband scattering \cite{Ferreira89}) as 
\begin{eqnarray}\label{HFSQ}
H_{Fr}&=&\sqrt{\frac{\hbar \omega_{LO}e^2}{2\epsilon_0 \epsilon_{\rho}SL_{BK}}}\sum_{\mathbf{q},q_z} 
\frac{ F(q_z)}{\sqrt{q^2+q_z^2}}\\&&
(\dc{q}{q_z}+\da{-q}{-q_z}) ( \cc{1}{k}\ca{2}{k+q}+\cc{2}{k-q}\ca{1}{k}),
\nonumber
\end{eqnarray}
where we have defined 
\begin{eqnarray}
\label{Fdef}
F(q)=\int dz  \chi_1(z)\chi_2(z)e^{-iq z}.
\end{eqnarray}

From \Eq{HFSQ} we see that, due to the three dimensional character of the LO-phonons
\cite{Stroscio01},  each electronic transition couples to multiple phonon modes, indexed by different values of the wavevector along the growth direction.
It is thus convenient to introduce second quantized operators corresponding to the 
particular linear superpositions of phonon modes that are coupled to electronic transitions
\begin{eqnarray}
\label{Dop}
\rc{q}&=&\frac{1}{\sqrt{A}}\sum_{q_z} \frac{F(q_z)\dc{q}{q_z}}{\sqrt{q^2+q_z^2}},\nonumber\\
\ra{q}&=&\frac{1}{\sqrt{A}}\sum_{q_z} \frac{\bar{F}(q_z)\da{q}{q_z}}{\sqrt{q^2+q_z^2}},
\end{eqnarray}
whose spatial wavefunctions along the $z$ axis are
\begin{eqnarray}
\label{phi}
\varphi_{q}(z)&=&\frac{1}{\sqrt{AL_{BK}}}\sum_{q_z} \frac{F(q_z)e^{iq_z z}}{\sqrt{q^2+q_z^2}}.
\end{eqnarray}
From Eqs. (\ref{phi}) and (\ref{Fdef}) we see that the intersubband transitions naturally couple to phonon modes localized inside the quantum well (it is easy to verify that $\varphi_{q}(z)$ vanishes to the first order in $q$ if $z$ is outside the common support of $\chi_1$ and $\chi_2$).  

The normalization factor $A$ can be fixed imposing bosonic commutation relations for the $\rc{q}$ operators
\begin{eqnarray}
\label{commr}
\comm{\ra{q}}{\rc{q'}}&=&\frac{1}{A}\sum_{q_z,q_z'} \frac{F(q_z)F(-q_z')}{\sqrt{(q^2+q_z^2)(q'^2+{q_z'}^{2})}}
 \comm{\da{q}{q_z}}{\dc{q'}{q'_z}}\nonumber
\\&=&\frac{\delta_{\mathbf{q,q'}}L_{BK}I(q)}{2Aq},
\end{eqnarray}
and thus
\begin{eqnarray}
A&=&\frac{L_{BK}}{2}\frac{I(q)}{q},
\end{eqnarray}
where we have defined 
\begin{eqnarray}
\label{Iq}
I(q)=\int dz dz' \chi_1(z)\chi_2(z)  \chi_2(z')\chi_1(z') e^{-q\lvert z-z' \lvert}.
 \end{eqnarray}

We can thus write Hamiltonians in \Eq{Hfreed} and \Eq{HFSQ} in terms of coherent $\rc{q}$ and $\bc{q}$ operators as
\begin{eqnarray}
\label{HFint}
H_{ph}&=&\sum_{\mathbf{q}}\hbar\omega_{LO} \rc{q}\ra{q},
\end{eqnarray}
and
\begin{eqnarray}
\label{Hint}
H_{Fr}&=&\sum_{\mathbf{q}}
\sqrt{N_{2DEG} \hbar \omega_{LO}\frac{e^2}{4\epsilon_0\epsilon_{\rho}}\frac{I(q)}{q}}(\bc{q}+\ba{-q})(\ra{q}+\rc{-q}),
\nonumber
\end{eqnarray}
where
\begin{eqnarray}
N_{2DEG}=\frac{N}{S}, 
\end{eqnarray} 
is the density of the two dimensional electron gas.
In order to pass from \Eq{Hfreed} to \Eq{HFint}, we are neglecting higher order phonon modes confined inside the quantum well. This is justified by the fact that we limit ourselves to long-wavelength modes.

\subsection{Superradiant Electron-Electron interaction}
 
\begin{figure}[t!]
\begin{center}
\includegraphics[width=8.5cm]{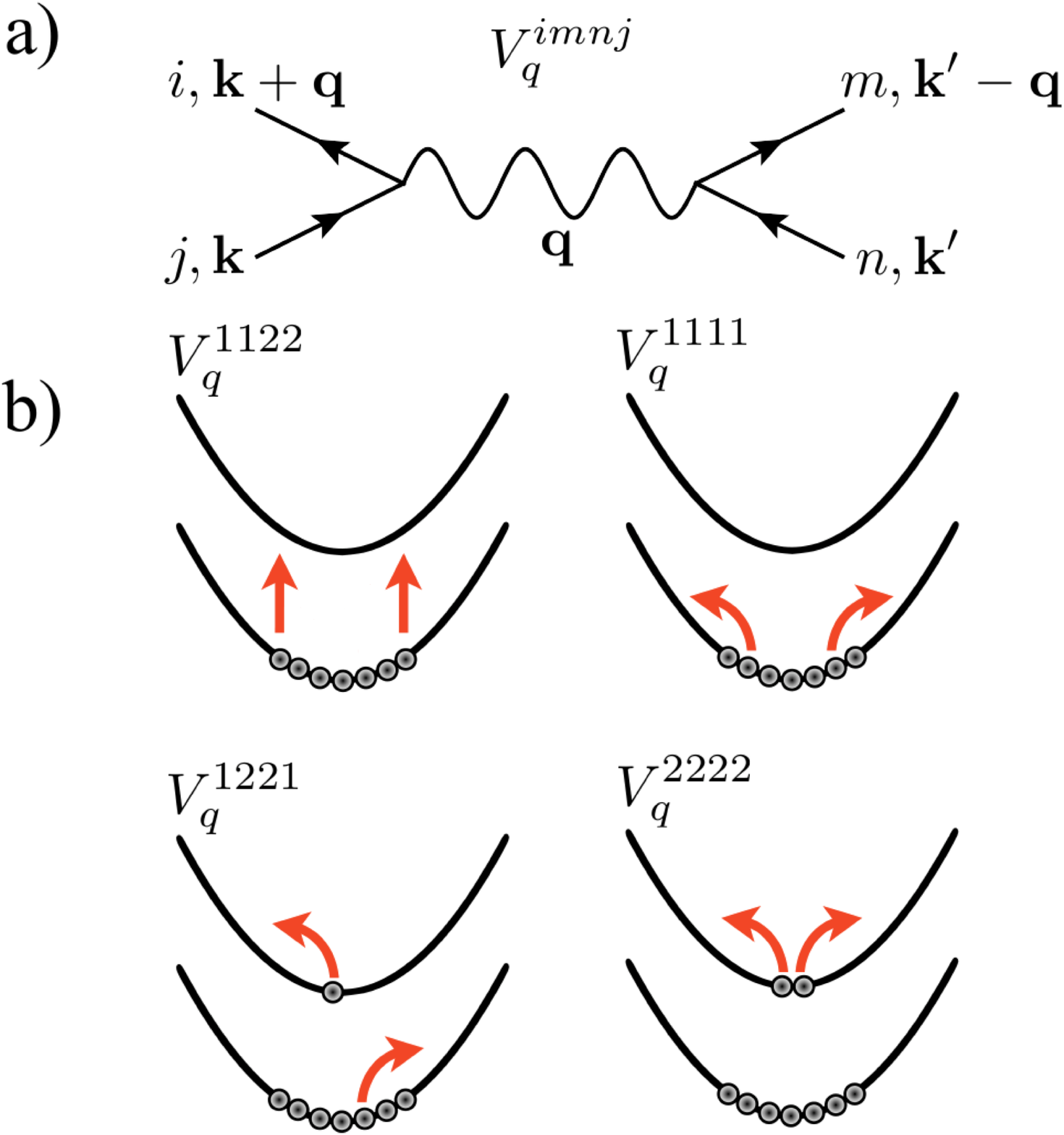}
\caption{ \label{CImage} a) Index convention of the matrix element $V_{q}^{imnj}$. Two electrons in subbands $j$ and $n$, with momenta $\mathbf{k}$
and $\mathbf{k'}$ are scattered into 
subbands $i$ and $m$,  with momenta $\mathbf{k+q}$
and $\mathbf{k'-q}$ respectively. b)  Graphical representation of the four qualitatively different kinds of scattering processes from \Eq{elem}.}
\end{center}
\end{figure}

In order to treat the Coulomb electron-electron interaction we start by the second quantized form of the Hamiltonian describing the Coulomb interaction \cite{Nikonov97} (see Fig. \ref{CImage} (a) for a graphical representation of the interaction coefficients) 
\begin{eqnarray}
\label{Hc1}
H_c=\frac{1}{2} \sum_{
i,j,m,n=1,2} \sum_{\mathbf{q, k, k'}} V^{imnj}_q \cc{i}{k+q} \cc{m}{k'-q} \ca{n}{k'} \ca{j}{k},\quad
\end{eqnarray}
where 
\begin{eqnarray}
\label{V}
V^{imnj}_q&=&\frac{e^2}{2\epsilon_0\epsilon_{\infty} q}\int dz dz' \chi_i(z)\chi_j(z)  \chi_m(z')\chi_n(z') e^{-q\lvert z-z' \lvert},\nonumber \\
\end{eqnarray}
is the two dimensional Coulomb matrix element.
It is important to notice that in \Eq{V} we used the high frequency dielectric constant $\epsilon_{\infty}$ instead of the static one. This is due to the fact that $\epsilon_{s}$ includes the effect of the coupling to LO-phonons, that are already treated exactly in the Hamiltonian.  

Due to the symmetry of the wavefunctions, a certain number of matrix elements in \Eq{V} can be seen to be zero, in particular all the matrix elements with an odd number of $1$ and $2$ indices
\begin{eqnarray}
\label{elem0}
V^{1112}_q&=& V^{1121}_q= V^{1211}_q= V^{2111}_q=0,\\
V^{2111}_q&=& V^{2212}_q= V^{2122}_q= V^{1222}_q=0.\nonumber 
\end{eqnarray}
The other elements can be evaluated as
\begin{eqnarray}
\label{elem}
V^{1122}_q&=& V^{1212}_q= V^{2121}_q= V^{2211}_q=\frac{e^2 I(q)}{2\epsilon_0\epsilon_{\infty} q},\\
V^{1221}_q&=& V^{2112}_q= \frac{e^2}{2\epsilon_0\epsilon_{\infty} q} \int dz dz' \chi^2_1(z)\chi^2_2(z') e^{-
 q\lvert z-z' \lvert}\nonumber,\\
 V^{1111}_q&=& \frac{e^2}{2\epsilon_0\epsilon_{\infty} q} \int dz dz' \chi^2_1(z)\chi^2_1(z') e^{-
 q\lvert z-z' \lvert}\nonumber,\\
V^{2222}_q&=&  \frac{e^2}{2\epsilon_0\epsilon_{\infty} q} \int dz dz' \chi^2_2(z)\chi^2_2(z') e^{-
 q\lvert z-z' \lvert}\nonumber,
\end{eqnarray}
where $I(q)$, defined in \Eq{Iq}, is the same integral we encountered studying the  
electron-phonon Fr\"ohlich interaction.  

The four distinct nonzero possible values of the matrix elements correspond to different kinds of scattering processes. In Fig. \ref{CImage} (b) a graphical representation for each of these processes is shown.

It is important at this point to notice a major difference between
the elements in the first  line of \Eq{elem} and the others. 
The elements in the first line (upper left subpanel in Fig. \ref{CImage} (b)) represent intersubband excitations: each electron is scattered from one subband to the other.
Such processes, responsible for the depolarization shift \cite{Nikonov97}, describe a superradiant process in the sense defined above, that is, at least for small values of $\mathbf{q}$, a great number of electrons can coherently undergo the same transition, approximately at the same energy. 
This is not the case for the interactions described in the other lines of \Eq{elem}, that instead describe intrasubband excitations that, either due to Pauli blocking or to the non-flat energy dispersions, involve only few electrons (see Fig. \ref{Cohex} for a graphical visualization of this crucial point). 

Our previous discussion on superradiant processes thus implies that the terms in the first line of \Eq{elem} strongly dominate over the others due to their superradiant enhancement. For this reason we have to treat them exactly in an Hamiltonian formalism, while we can  limit ourselves to treat the others within a perturbative approach.

Here we will thus construct an exact, Hamiltonian approach, to treat the effect of the depolarization shift terms, neglecting the others. We will analyze later the effect of the intrasubband terms.

Let us start to rewrite the depolarization shift part of  \Eq{Hc1} in a more useful form
\begin{widetext}
\begin{eqnarray}
\label{HCnew}
H_c&=& \sum_{\mathbf{q, k, k'}} \frac{e^2 I(q)}{4\epsilon_0\epsilon_{\infty} q}\Big(
\cc{1}{k+q} \cc{1}{k'-q} \ca{2}{k'} \ca{2}{k} +
\cc{1}{k+q} \cc{2}{k'-q} \ca{1}{k'} \ca{2}{k} + 
\cc{2}{k+q} \cc{1}{k'-q} \ca{2}{k'} \ca{1}{k} +
\cc{2}{k+q} \cc{2}{k'-q} \ca{1}{k'} \ca{1}{k}\Big)\nonumber \\
 &=&
\sum_{\mathbf{q, k, k'}}  \frac{e^2 I(q)}{4\epsilon_0\epsilon_{\infty} q}\Big(
\cc{1}{k+q} \ca{2}{k} \cc{1}{k'-q} \ca{2}{k'} +
\cc{1}{k+q} \ca{2}{k} \cc{2}{k'-q} \ca{1}{k'}+
\cc{2}{k+q} \ca{1}{k} \cc{1}{k'-q} \ca{2}{k'} +
\cc{2}{k+q} \ca{1}{k}\cc{2}{k'-q} \ca{1}{k'} \Big)\nonumber\\
&&+\sum_{\mathbf{q}} \frac{Ne^2 I(q)}{4\epsilon_0\epsilon_{\infty} q}.
\end{eqnarray}
\end{widetext}
We see from \Eq{HCnew} that, thanks to its collective, superradiant nature, the depolarization shift can be naturally written in terms of the bosonic intersubband excitations defined in \Eq{bdef} as 
\begin{eqnarray}
\label{Hc2}
H_c= \sum_{\mathbf{q}} N_{2DEG} \frac{e^2}{4\epsilon_0 \epsilon_{\infty}} \frac{I(q)}{q} (\bc{q}+\ba{-q}) (\bc{-q}+\ba{q}),
\end{eqnarray}
 where we have neglected the last constant term, that simply shifts the ground state energy.

 \subsection{Residual Electron-Electron interaction}
 
We treated exactly the depolarization shift terms of the Coulomb interaction in a bosonic excitation formalism.
Moreover, we showed how the terms other that the ones responsible for the depolarization shift are strongly suppressed, due to their lack of collective enhancement and can thus be treated perturbatively.

Here we will study the perturbative effect of such residual Coulomb contributions, due to the intrasubband terms in the last three lines of \Eq{elem} (schematized in the last three panels of Fig. \ref{CImage} (b)). 

An important result due to Lee and Galbraith \cite{Lee99,Lee00},  is that such intrasubband terms do not contribute to the screening of the intersubband ones at the level of the random phase approximation (RPA).
This can be seen writing the Dyson equation for the dynamically screened Coulomb potential  \cite{Sotirelis93} $\mathcal{V}_q(\omega)$
\begin{eqnarray}
\label{DysonC}
\mathcal{V}_q^{imnj}(\omega)=V_q^{imnj}+\sum_{rs} V_q^{irsj} \Pi^{sr}_q(\omega) \mathcal{V}_q^{smnr}(\omega),
\end{eqnarray}
where $\Pi^{sr}_q(\omega)$ is the RPA polarization function.
In the case of an intersubband contribution (e.g., $\mathcal{V}_q^{1122}$), Eqs. (\ref{elem0}) and (\ref{elem}) imply that
\begin{eqnarray}
\label{Dysonisbt}
\mathcal{V}_q^{1122}(\omega)&=&V_q^{1122}+\sum_{rs} V_q^{1rs2} \Pi^{sr}_q(\omega) \mathcal{V}_q^{s12r}(\omega) \\
&=&V_q^{1122}+\sum_{r \neq s} V_q^{1rs2} \Pi^{sr}_q(\omega) \mathcal{V}_q^{s12r}(\omega)  \nonumber\\
&=&V_q^{1122}+V_q^{1122} (\Pi^{12}_q(\omega)  +\Pi^{21}_q(\omega) )  \mathcal{V}_q^{1122}(\omega)\nonumber.
\end{eqnarray}
We have thus
\begin{eqnarray}
\label{Dysonscreening}
\mathcal{V}_q^{1122}(\omega)&=&\frac{V_q^{1122}}{1-V_q^{1122} (\Pi^{12}_q(\omega)  +\Pi^{21}_q(\omega) ) },
\end{eqnarray}
from which we see that the intrasubband Coulomb terms ($V_q^{1111}$,$V_q^{2222}$,
$V_q^{1221}$ and $V_q^{2112}$) do not intervene in the renormalization of the intersubband terms.

An analogous reasoning can be done for the phonon-electron interaction. Calling $M_{q,q_z}$ and $\mathcal{M}_{q,q_z}(\omega)$ the bare and screened version of the potential defined in \Eq{HFSQ}, we have the Dyson equation
\begin{eqnarray}
 \label{DysonF}
\mathcal{M}_{q,q_z}(\omega)&=&M_{q,q_z}+\sum_{rs}  V_q^{1mn2}  \Pi^{sr}_q(\omega) \mathcal{M}_{q,q_z}(\omega)\\
&=&M_{q,q_z}+ V_q^{1122}(  \Pi^{12}_q(\omega)+ \Pi^{21}_q(\omega) )\mathcal{M}_{q,q_z}(\omega)\nonumber,
\end{eqnarray}
and thus the formula for the screened potential is
\begin{eqnarray}
 \label{Dysonscreeningphoon}
\mathcal{M}_{q,q_z}(\omega)&=&
\frac{M_{q,q_z}}{1-V_q^{1122}(  \Pi^{12}_q(\omega)+ \Pi^{21}_q(\omega) )}.
\end{eqnarray}
Being the RPA screening only due to terms that are exactly treated in the Hamiltonian, we can thus neglect the screening due to the two dimensional electron gas. 
 
 \subsection{Hopfield-Bogoliubov Hamiltonian}
 
Putting together Eqs. (\ref{Hfreeb}), (\ref{Hint}) and (\ref{Hc2}) we arrive to the full Hamiltonian for the intersubband transitions-LO-phonons system
\begin{eqnarray}
\label{Hfull0}
H&=&\sum_{\mathbf{q}} \hbar\omega_{12}\bc{q}\ba{q}
+\hbar\omega_{LO} \rc{q}\ra{q}\\&& \nonumber
+\sqrt{N_{2DEG} \hbar \omega_{LO}\frac{e^2}{4\epsilon_0\epsilon_{\rho}}\frac{I(q)}{q}}(\bc{q}+\ba{-q})(\rc{-q}+\ra{q})
\\&& \nonumber +N_{2DEG} \frac{e^2}{4\epsilon_0 \epsilon_{\infty}} \frac{I(q)}{q} (\bc{q}+\ba{-q}) (\bc{-q}+\ba{q}).
\end{eqnarray}
The Hamiltonian in \Eq{Hfull0} can be rewritten in a more compact form by introducing the intersubband transitions-LO-phonons coupling coefficient $\Omega$ and the Coulomb coefficient $D$
 \begin{eqnarray}
\label{Omega}
\Omega&=&\sqrt{N_{2DEG} \omega_{LO}\frac{e^2}{4\epsilon_0\epsilon_{\rho} \hbar}\frac{I(q)}{q}},\\
D&=&N_{2DEG} \frac{e^2}{4\epsilon_0 \epsilon_{\infty}} \frac{I(q)}{q},\nonumber
\end{eqnarray}
where we have dropped the dependences over the wavevector as we are interested in the long wavelength limit (from \Eq{Iq} we can verify that $\lim_{q\rightarrow 0}\frac{I(q)}{q}$ tends to a constant value).

Using \Eq{Omega}, \Eq{Hfull0} can be written as
 \begin{eqnarray}
\label{Hfull}
H&=&\hbar \sum_{\mathbf{q}} \omega_{12}\bc{q}\ba{q}
+\omega_{LO} \rc{q}\ra{q}
+\Omega(\bc{q}+\ba{-q})(\rc{-q}+\ra{q})
 \nonumber \\&& +D(\bc{q}+\ba{-q}) (\bc{-q}+\ba{q}),
\end{eqnarray} 
 that can be cast in matrix form as
\begin{equation}
H=\frac{\hbar}{2}\, \sum_{\mathbf{q}}\vhd_{\mathbf{q}}\,\eta\,\Mc_q\,\vh_{\mathbf{q}}, 
\label{Hamilt}
\end{equation}
where the column vector of operators $\vh_{\mathbf{q}}$ is defined
as 
\begin{eqnarray}
\vh_{\mathbf{q}}=
\lbrack
\ba{q},\ra{q},\bc{-q},\rc{-q}
\rbrack^T,
\end{eqnarray}
$\eta$ is the diagonal metric
 \begin{eqnarray}
 \eta=\textrm{diag}[1,1,-1,-1],
 \end{eqnarray}
 and the Hopfield-Bogoliubov\cite{Hopfield58} matrix $\Mc_q$ is defined as
\begin{equation}
\label{HBM}
\Mc_q = \left(
\begin{array}{cccc}
\omega_{12}+2D &  \Omega&
2D&\Omega\\
\Omega&\omega_{LO} &\Omega&0
\\
-2 D&-\Omega
& -\omega_{12}-2D&-\Omega\\
-\Omega&0&-\Omega&
-\omega_{LO}
\\
\end{array} \right ).
\end{equation}

Diagonalizing the matrix in \Eq{HBM} will yield the frequencies of the normal modes of the system $\omega_{\pm}$, that are usually called polarons \cite{Hameau99,Verzelen02}. In our case we will name them more properly intersubband polarons, because the electronic part of the mixed excitations is an intersubband transition.
Hamiltonian in \Eq{Hfull} can thus be put in the diagonal form
 \begin{eqnarray}
\label{HD}
H&=&\sum_{j=\pm,\mathbf{q}} \hbar\omega_{j}p^{\dagger}_{j,\mathbf{q}}p_{j,\mathbf{q}}+E_{\Delta},
\end{eqnarray}
where the $p_{j,\mathbf{q}}$ are the annihilation operators for the two polaronic branches, given by a linear superposition of $\ba{q}$, $\ra{q}$, $\bc{-q}$ and $\rc{-q}$ operators and $E_{\Delta}$ is the energy of the new ground state relative to the one of the uncoupled system.

\subsection{Coupled Ground State}

The coupling between the intersubband transitions and the LO-phonons does not modify only the system's resonances but it also qualitatively modifies the nature of its ground state.
It is easy to verify that, if $\ket{0}$ is the ground state for the uncoupled phonons and intersubband excitations, defined in the usual way as
\begin{eqnarray*}
\ba{q}\ket{0}&=&\ra{q}\ket{0}=0,
\end{eqnarray*}
then 
\begin{eqnarray*}
p_{j,\mathbf{q}}\ket{0}&\neq&0,
\end{eqnarray*}
that is, $\ket{0}$ is not the ground state for the coupled system. 
The real ground state of the Hopfield matrix in \Eq{HBM}, that has been thoroughly studied in Ref. [\onlinecite{Ciuti05}], has the form of a two modes squeezed vacuum.

Still, thanks to the bosonicity of the system, such new ground state does not influence the response of the system, that can be described as a gas of free bosonic excitations (from \Eq{HD}).
A notable exception is the case in which the parameters of the system are nonadiabatically modulated in time. In this case the sudden change in the ground state \cite{Gunter09} can have observable effects,
like the emission of quantum vacuum radiation \cite{DeLiberato07,DeLiberato09}.

It is also interesting to notice that, from \Eq{Omega}, we can write the Coulomb coefficient $D$ as
 \begin{eqnarray}
\label{DO}
D&=&\frac{\Omega^2}{\omega_{LO}}\frac{\epsilon_{\rho}}{\epsilon_{\infty}}\geq \frac{\Omega^2}{\omega_{LO}}.
\end{eqnarray}
As it has recently been shown in Ref. \lbrack\onlinecite{Nataf10}\rbrack, \Eq{DO} implies that the ground state of the system will not undergo a Dicke phase transition, regardless of the strength of the coupling.

\subsection{Multiple Quantum Wells}

Until now we considered the case of a single quantum well. This choice has been motivated by the fact that, as we will show, the presence of multiple wells does not modify our results.

Given that we are considering rather large quantum wells (in order for the transition to be resonant with the LO-phonon mode), the optical phonon spectrum is not modified \cite{Stroscio01} and the optical phonon modes we consider are confined in each quantum well.

This is a rather important difference between the intersubband polaron case we consider in this paper and the physics of intersubband polaritons.
For intersubband polaritons,  the electromagnetic mode coupled to the intersubband transitions extends over all the structure. It thus couples to all the electrons, regardless of the quantum well they are in.
This means that  the only meaningful parameter for intersubband polaritons is the total density of electrons, and the light-matter coupling thus scales as  $\sqrt{n_{QW}N_{2DEG}}$, where $n_{QW}$ is the number of quantum wells inside the microcavity.

In the present case instead, being the phonon modes confined inside each quantum well, electrons in different wells are completely decoupled. This can also
be inferred from the coupling integral in \Eq{V}. This integral does vanish, at least in the long wavelength limit (first order in $q$), if the wavefunctions for the two integration variables $z$ and $z'$ do not have a common support, i.e., if the two interacting electrons are in different quantum wells. 

This means that, contrary to the intersubband polariton case,  the intersubband polaron interaction scales only as $\sqrt{N_{2DEG}}$ and growing multiple quantum wells in the same sample will not increase the coupling.

\section{Results}
 
 \label{NumRes}
 
 In order to obtain some numerical predictions from Hamiltonian in \Eq{Hfull}, we need to fix a few parameters concerning the material and the quantum well. 
 
For sake of simplicity we will consider the quantum well to be correctly approximated by a rectangular, infinite potential well of length $L_{QW}$. We thus have
 \begin{eqnarray}
\hbar \omega_{12}&=&\frac{3\hbar^2 \pi^2}{2m^*L_{QW}^2},
\end{eqnarray} 
and the electronic and phononic modes profiles are given by
 \begin{eqnarray}
 \label{inf}
\chi_1(z)&=&\sqrt{\frac{2}{L_{QW}}}\sin(\frac{\pi z}{L_{QW}}),\\
\chi_2(z)&=&\sqrt{\frac{2}{L_{QW}}}\sin(\frac{ 2\pi z}{L_{QW}}),\nonumber \\
\varphi_{0}(z)&=&\sqrt{\frac{16}{5L_{QW}}}\sin^3(\frac{ \pi z}{L_{QW}}),\nonumber
\end{eqnarray} 
inside the well and zero outside.
As explained in Sec. \ref{Theory}, we see here explicitly that the intersubband transitions couple to a linear superposition of phonon modes that is localized inside the quantum well (the cubic sinus in the third line of \Eq{inf} comes from the integral of the first two, as can be verified performing the integral in \Eq{phi}).

Inserting \Eq{inf} into \Eq{Iq} and performing the integral we have
 \begin{eqnarray}
\label{lim}
\lim_{q\rightarrow 0} I(q) \rightarrow \frac{10}{9\pi^2}qL_{QW}.
\end{eqnarray}

\begin{figure}[t!]
\begin{center}
\includegraphics[width=9cm]{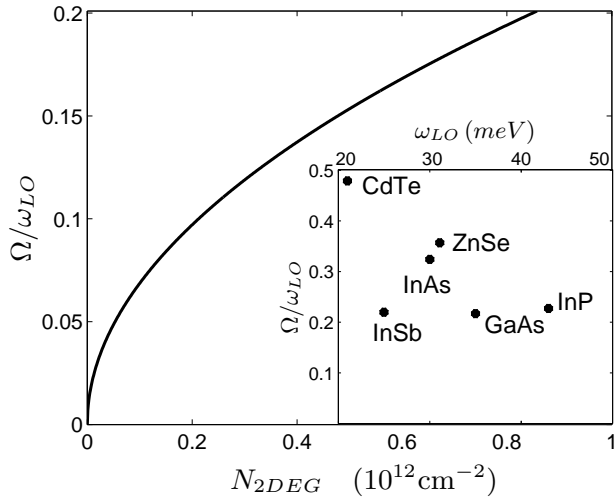}
\caption{ \label{coupling} Normalized coupling $\frac{\Omega}{\omega_{LO}}$ in GaAs as a function of the doping density $N_{2DEG}$. Inset:
the same quantity as a function of the LO-phonon frequency $\omega_{LO}$ for different materials, for $N_{2DEG}=10^{12}$cm$^{-2}$.}
\end{center}
\end{figure}

In Fig. \ref{coupling} we plot the normalized coupling $\frac{\Omega}{\omega_{LO}}$ as a function of the density of the two dimensional electron gas, for a GaAs quantum well. 
In the inset of Fig. \ref{coupling} we instead present a comparison of the values of $\frac{\Omega}{\omega_{LO}}$,  at room temperature, for different semiconductors of the III-V and II-VI groups\cite{Materials}, as a function of the respective LO-phonon energies, for a reference doping $N_{2DEG}=10^{12}$cm$^{-2}$.  

\begin{figure}[t!]
\begin{center}
\includegraphics[width=9cm]{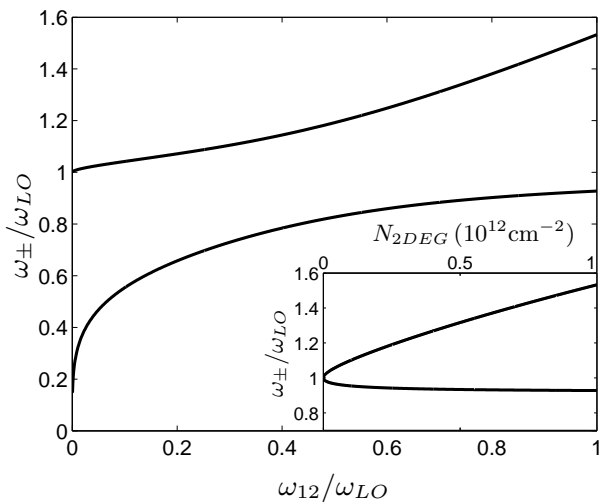}
\caption{ \label{dispersion} Intersubband polaron frequencies $\omega_{\pm}$ as a function of the intersubband frequency $\omega_{12}$ in GaAs for $N_{2DEG}=10^{12} cm^{-2}$. Inset: the same quantity as a function of the doping for  $\omega_{12}=\omega_{LO}$.}
\end{center}
\end{figure}

In Fig. \ref{dispersion} there is a plot of the intersubband polaron frequencies $\omega_{\pm}$ 
as a function of the intersubband frequency $\omega_{12}$, in GaAs, for $N_{2DEG}=10^{12}$cm$^{-2}$. Notice that, due to the effect of Coulomb interaction, the resonant anticrossing is not at $\omega_{12}=\omega_{LO}$ but at a lower frequency. 
In the inset of the same figure we plot the same quantity as a function of the electron density. The length $L_{QW}$ has been chosen in this case to have the two uncoupled modes at resonance
($\omega_{12}=\omega_{LO}$, that is $L_{QW}\simeq 23$nm).
 
It is clear from the figures that intersubband polarons are not only strongly coupled, having coupling constants much larger than their linewidth (usual linewidths being not bigger than a few meV), but they
are indeed in the ultrastrong coupling regime, with values of the normalized coupling $\frac{\Omega}{\omega_{LO}}$ comparable or larger than the best ones reported in the literature.
For physically realizable levels of doping,  coupling values of a few tenths of the bare frequency of the excitation $\omega_{LO}$ are predicted in GaAs, and it seems that values much larger can be obtained using more polar materials.
The reason of such large coupling can be found in the superradiant nature of intersubband excitations and in the natural confinement of the phonons inside the quantum well, that gives an extremely small mode volume, when compared with what can be obtained with photonic microcavities. 
 
The consequences of our results can be multiple, both for fundamental and applied research. On the fundamental side, intersubband polarons could become a new laboratory to test quantum vacuum physics, typical of the ultrastrong coupling regime \cite{DeLiberato07}. On the applied side our theory can be naturally exploited in the study of quantum cascade lasers working in or near the Restrahlen band. It can, for example, help explaining the anticrossing observed in Ref. \onlinecite{Castellano11}, near the LO-phonon frequency. Moreover the capability to strongly modify the LO-phonon spectrum could have an impact on the performances of optoelectronic devices, as the electron-LO-phonon scattering rate determines the lifetime of carriers in excited subbands \cite{Ferreira89}.

\psfrag{XO}[Bc][B][1.2]{$\begin{array}{c}\\  \quad N_{2DEG} \quad(10^{12}$cm$^{-2}) \end{array}$}
\psfrag{xO}[Bc][B][1]{$\omega_{LO} \, (meV)$}
\psfrag{YO}[B][B][1.2]{$\Omega/\omega_{LO}$}
\psfrag{yO}[B][B][1]{$\Omega/\omega_{LO}$}

\psfrag{XD}[Bc][B][1.2]{$\begin{array}{c}\\ \omega_{12}/\omega_{LO}  \end{array}$}
\psfrag{xD}[Bc][B][1]{$\,\quad N_{2DEG}\,(10^{12}$cm$^{-2})$}
\psfrag{YD}[B][B][1.2]{$\omega_{\pm}/\omega_{LO}$}
\psfrag{yD}[B][B][1]{$\omega_{\pm}/\omega_{LO}$}

\section{Comparison with dielectric function theory}
\label{Appendix}

In the previous Sections we have developed a detailed microscopic theory for the intersubband transitions coupled to LO-phonons.
Here we will compare the dispersions obtained from the microscopic theory with the ones obtained with an homogeneous dielectric function theory, as the one used in Ref. \lbrack\onlinecite{Liu03}\rbrack.

The propagation of an electromagnetic wave in a dispersive, homogeneous medium obeys the equation
\begin{eqnarray}
\label{divD}
\text{div}\lbrack D(\omega) \rbrack=\text{div}\lbrack \epsilon(\omega) E(\omega)\rbrack=0.
\end{eqnarray}
This implies that it is possible to have propagating longitudinal waves, like polarons, only at frequencies for which 
\begin{eqnarray}
\label{epszero}
\Re\lbrack \epsilon(\omega)\rbrack&=&0,
\end{eqnarray}
where $\Re$ indicates the real part.

The $z$ component of the dielectric function of an homogeneous medium filled with quantum wells is given by \cite{Liu03,Todorov10} 
\begin{eqnarray}
\label{Aeps}
\epsilon(\omega)&=&\epsilon_{\infty}\frac{\omega^2-\omega_{LO}^2}{\omega^2-\omega_{TO}^2+i\omega 0^+}-\epsilon_{\infty}\frac{\omega_{P}^2}{\omega^2-\omega_{12}^2+i\omega 0^+},\nonumber \\
\end{eqnarray}
where
\begin{eqnarray}
\label{omegap}
\omega^2_P=\frac{2\omega_{12}d_{12}^2N_{2DEG}}{\hbar \epsilon_0 \epsilon_{\infty} L_{QW}},
\end{eqnarray}
is the plasma frequency of the two dimensional electron gas and $d_{12}$ is the intersubband dipole
\begin{eqnarray}
\label{d12}
d_{12}=e\int dz \chi_1(z)z\chi_2(z).
\end{eqnarray}
The equation 
\begin{eqnarray}
\label{Aepszero}
\Re\lbrack \epsilon(\omega)\rbrack&=&0,
\end{eqnarray}
thus reads
\begin{eqnarray}
\label{Anumeps}
\omega^4-\omega^2(\omega_{LO}^2+\omega_{12}^2+\omega_P^2)+\omega_{LO}^2\omega_{12}^2+\omega_{TO}^2\omega_{P}^2&=&0.\nonumber \\
\end{eqnarray}

As \Eq{Aeps} neglects both the dielectric response in the  $x-y$ plane and the non-homogeneity in the $z$ direction, in order to recover the same result from our microscopic approach, we will have to consider only phonon modes with $\mathbf{q}=0$ and  $q_z\rightarrow 0$.
From Eqs. (\ref{Fdef}) and (\ref{d12}) we thus have
\begin{eqnarray}
\label{AFdef}
\frac{F(q_z)}{\sqrt{q^2+q_z^2}}\rightarrow -i\frac{d_{12}}{e}.
\end{eqnarray}

Following exactly the same procedure of Sec. \ref{Theory}, but with the $F(q)$ defined in \Eq{AFdef} and considering only the $q_z\rightarrow 0$ mode, we get
\begin{eqnarray}
\label{AOmega}
\Omega&=&\sqrt{\frac{N_{2DEG}\omega_{LO}d_{12}^2}{2\epsilon_0\epsilon_{\rho}L_{QW} \hbar}},
\end{eqnarray}
and thus, from \Eq{DO}
\begin{eqnarray}
\label{AD}
D&=&\frac{N_{2DEG}d_{12}^2}{2\epsilon_0\epsilon_{\infty}L_{QW} \hbar}.
\end{eqnarray}

In order to obtain the polaronic eigenfrequencies we have to diagonalize the matrix in \Eq{HBM} using the coupling coefficients for the homogeneous limit defined in Eqs. (\ref{AOmega}) and (\ref{AD}). 
We thus obtain the secular equation
\begin{eqnarray}
\label{Anumdiag}
&&\omega^4-\omega^2(\omega_{LO}^2+\omega_{12}^2+4D\omega_{12})\\&&+\omega_{LO}^2\omega_{12}^2+4D\omega_{12}\omega_{LO}^2-4\Omega^2\omega_{12}\omega_{LO}=0,\nonumber 
\end{eqnarray}
that, using Eqs. (\ref{AD}) and (\ref{omegap}) can be put into the form
\begin{eqnarray}
\label{Anumdiag2}
\omega^4-\omega^2(\omega_{LO}^2+\omega_{12}^2+\omega_{P}^2)+\omega_{LO}^2\omega_{12}^2+\omega_{P}^2\omega_{LO}^2\frac{\epsilon_{\infty}}{\epsilon_{s}}=0.\nonumber \\
\end{eqnarray}
Equating the coefficients of Eqs. (\ref{Anumeps}) and (\ref{Anumdiag2}), we obtain
\begin{eqnarray}
\label{LST}
\omega_{TO}^2&=&\omega_{LO}^2\frac{\epsilon_{\infty}}{\epsilon_{s}},
\end{eqnarray}
that is the well known Lyddane-Sachs-Teller relation \cite{Stroscio01}.
We have thus proved that the homogeneous version of our theory gives the same results as the homogeneous dielectric function approach.

It is anyway important to notice that the homogeneous limit in not exact, 
as a quantum well is, by definition, spatially inhomogeneous. Ignoring the higher $q_z$ modes leads to underestimate the intersubband dipole of a factor roughly equal to $\sqrt{2}$.

 \section{Conclusions}
 \label{Con}
In this paper we have developed a microscopic theory of intersubband polarons, mixed excitations resulting from the coupling between intersubband transitions in doped quantum wells and LO-phonons. We took into account the electron-electron Coulomb interaction and we were able to treat exactly the resulting depolarization shift.
We proved that intersubband polarons can be in the ultrastrong coupling regime, reaching extremely high values of the coupling constant.
We critically discussed the relevance of our results both for fundamental and applied research.

\section{Acknowledgments}
We would like to thanks  D. Hagenm\"uller, M. Zaluzny, P. Nataf,  L. Nguyen, J. Restrepo,  C. Sirtori  and Y. Todorov for useful discussions and comments.  C. C. is member of Institut Universitaire de France. We acknowledge support from the ANR grant
QPOL

\end{document}